\documentclass{tufte-handout}

\usepackage[T1]{fontenc}
\usepackage[utf8]{inputenc}
\usepackage{mathpazo}                   
\usepackage[scaled=0.90]{helvet}        
\usepackage{courier}                    

\setcitestyle{numbers,square}

\usepackage{amsmath,amssymb}
\usepackage{booktabs}
\usepackage{tabularx}
\usepackage{graphicx}
\graphicspath{
  {figures/}
  {../Figures/}
  {/Users/bozprog/Library/Mobile Documents/com~apple~CloudDocs/FIGURES/}
}
\usepackage{microtype}
\usepackage{hyperref}
\usepackage{xcolor}
\usepackage{tikz}
\usepackage{eso-pic}
\usepackage{currfile}
\usepackage{enumitem}

\hypersetup{
    colorlinks=true,
    linkcolor=darkgray,
    citecolor=darkgray,
    urlcolor=darkgray
}

\newcommand{\fito}{FITO}
\newcommand{\oae}{OAE}
\newcommand{\paxos}{\textsc{Paxos}}

\newcommand{\docversion}{v0.1}
\usepackage{datetime2}
\DTMsetdatestyle{iso}

\fancypagestyle{firstpage}{%
  \fancyhf{}%
  \fancyfoot[L]{\raisebox{-0.8in}{\tiny\texttt{\docversion\ -- \today}}}%
  \fancyfoot[R]{\raisebox{-0.8in}{\tiny\texttt{\currfilename}}}%
}
\newcommand{\showfilename}{\thispagestyle{firstpage}}

\newcommand{\placelogo}{%
  \AddToShipoutPictureBG*{%
    \AtPageUpperLeft{%
      \raisebox{-0.75in}{\hspace{\dimexpr\paperwidth-1.25in\relax}%
        \includegraphics[height=0.6in]{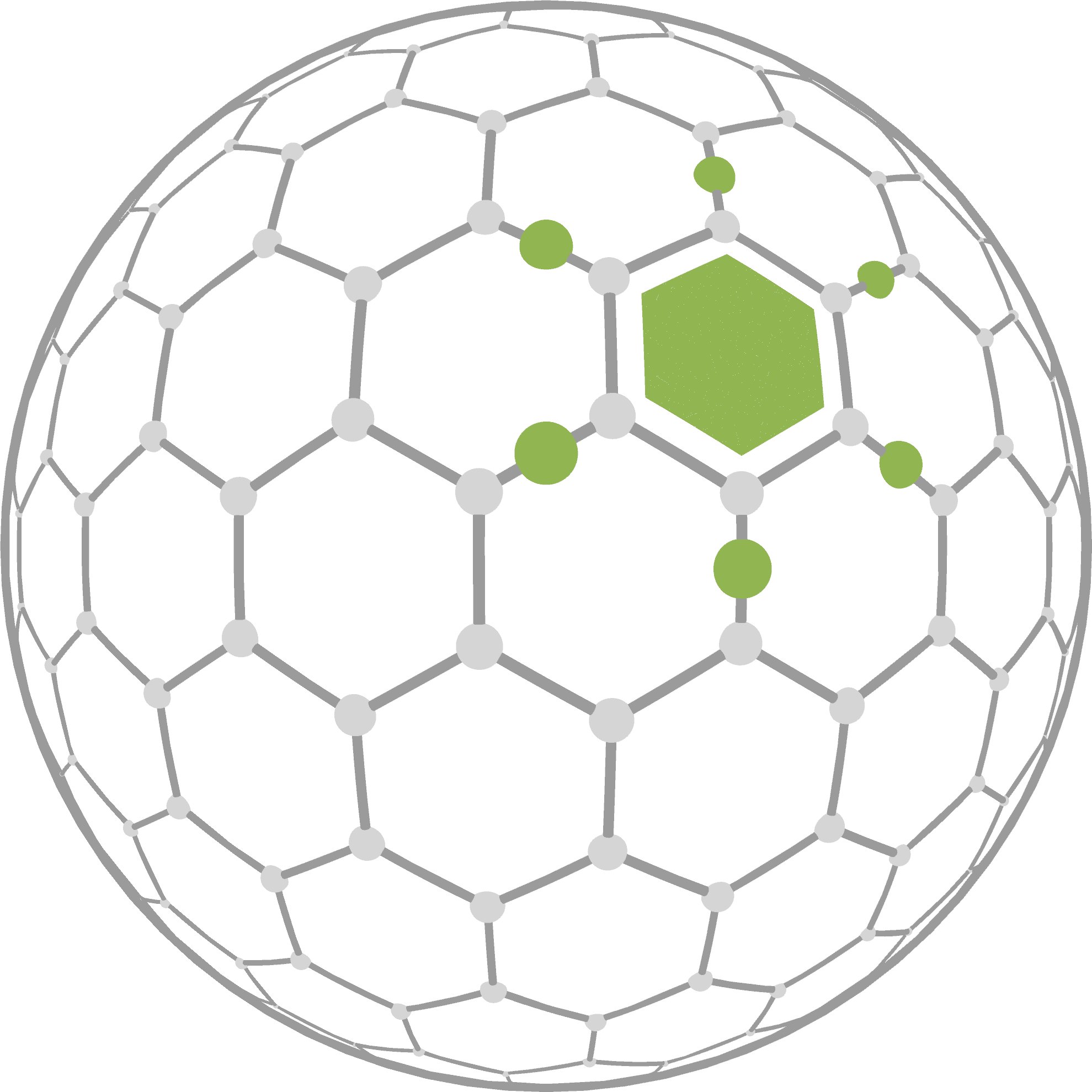}%
      }%
    }%
  }%
}

\title[Background and Intellectual Development]{Background and Intellectual Development\\\medskip
{\large Supplementary Material for the Category Mistake Papers}}

\author{Paul Borrill\thanks{DAEDAELUS. \texttt{paul@daedaelus.com}. ORCID: 0000-0002-7493-5189.}}

\date{March 2026}

\begin{document}

\maketitle
\showfilename
\placelogo

\begin{abstract}
\noindent This supplement documents the intellectual trajectory that led to the \emph{Category Mistake} framework and the \emph{Forward-In-Time-Only} (\fito{}) analysis presented in our recent arXiv papers. The ideas crystallized over fifteen years of research, conversation, and engineering practice---beginning with a 2014 Stanford EE380 lecture on the physics of time in computing, sharpened through a 2016 email exchange with Leslie Lamport following a Papers We Love presentation of his seminal 1978 paper, and matured through the development of Open Atomic Ethernet (\oae{}). This document traces the concept development from its origins in the physics of entanglement and background-free time, through the recognition that Lamport's ``happened-before'' relation embeds a category mistake, to the practical engineering consequences documented in ``Why iCloud Fails'' and ``What Distributed Computing Got Wrong.'' It is intended as archival supplementary material for future arXiv submission.
\end{abstract}

\section{Origins: The Physics of Time (2013--2014)}

The intellectual foundation of the Category Mistake framework begins not in computer science but in theoretical physics. The central insight---that \emph{time is change we can count}, not a background against which change is measured---emerged from engagement with three converging lines of physics research.

\marginnote{The phrase ``time is change we can count'' is Aristotelian in origin. Aristotle explains time in terms of change, not vice versa. See \emph{Time for Aristotle}, Notre Dame Philosophical Reviews.}

\subsection{Background-Free Physics}

Modern physics has progressively undermined the notion of a continuous temporal background. Newton's absolute time gave way to Einstein's relative time, which itself rests on Minkowski's assumption of a smooth four-dimensional manifold. By 2013, a critical mass of theoretical physicists---Arkani-Hamed, Maccone, Susskind, Maldacena, Van Raamsdonk, Carroll, Lloyd, Ooguri---had converged on a striking claim: spacetime is not fundamental but \emph{emergent}, arising from quantum entanglement.\citep{maccone2009,moreva2014,lloyd2014,vanraamsdonk2010}

\marginnote[-0.5cm]{Arkani-Hamed at Cornell: ``Space-time is doomed. What replaces it?'' Susskind and Maldacena: ER=EPR---wormholes and entanglement are the same phenomenon.}

The experimental illustration came from Moreva et al.\citep{moreva2014}, who demonstrated that an entangled photon pair exhibits temporal evolution to an internal observer while appearing static to an external one---exactly the Page-Wootters mechanism\citep{pagewootters1983} predicted in 1983. Time, in this picture, emerges from correlations between subsystems, not from a pre-existing background.

\subsection{The Subtime Conjecture}

These physics insights were synthesized into what became known as the \emph{Subtime Conjecture}, first articulated in the Simple Time Description document\citep{borrill2016simpletime} and presented at Stanford EE380 on April 16, 2014\citep{borrill2014stanford}:

\begin{enumerate}[nosep]
    \item Time is change that we can count.
    \item By change, we mean quantum mutual information.
    \item Mutual information is bipartite---a single bit shared between two entities.
    \item This single bit \emph{appears} to be ``in two places at the same time,'' but this appearance is an artifact of the assumption of a Newtonian or Minkowski background.
    \item An alternative model: information is physical and can be ``trapped in perpetuity'' between two elementary particles through perpetual exchange.
    \item Time may therefore go forward and backward at the local interaction level; each elementary bipartite system has ``its own'' time.
\end{enumerate}

\marginnote{The Subtime Conjecture draws directly on Wheeler-Feynman absorber theory: a photon is not ``emitted'' until it is ``absorbed.'' Causality is bilateral, not unidirectional.}

The conjecture carried two provocative corollaries: (a) entanglement is real and is the root of causality, and (b) quantum computing speedup---which is predicated on a Minkowski background against which ``durations'' can be measured---may be illusory. These remain open questions in physics, but the \emph{engineering} consequences proved immediate and practical.

\subsection{The Alice-and-Bob Universe}

The Simple Time Description\citep{borrill2016simpletime} introduced a thought experiment that became the conceptual backbone of the entire program. Imagine a universe containing only two hydrogen atoms---Alice and Bob---and a single photon of appropriate energy to be absorbed and re-emitted between them.

From Alice's local-observer-view (LOV), there are exactly two states: she has the photon, or she does not. From the photon's LOV, there are only two events: arrival at Alice and arrival at Bob. The proper time for a photon is zero; Alice and Bob are ``indistinguishable'' from the photon's perspective.

\marginnote{This thought experiment is the miniature version of what OAE implements at the link level: two endpoints exchanging a token, with no background time and no intermediate ``in flight'' ambiguity.}

The conclusion, deliberately paraphrasing Einstein's 1905 paper on the electrodynamics of moving bodies: ``The introduction of `spacetime' will prove to be superfluous inasmuch as the view here to be developed will not require a `continuous background' provided with special properties, such as memory of duration.''

This was not idle philosophizing. The Alice-and-Bob model directly informed the design of OAE's link protocol, where two endpoints exchange a causally significant token with no intermediate state of ambiguity.

\section{The Lamport Encounter (2016)}

The bridge from physics to distributed systems was constructed through direct engagement with Leslie Lamport's work, beginning with the July 14, 2016 Papers We Love talk at GitHub headquarters in San Francisco.\citep{borrill2016pwl}

\subsection{Papers We Love: San Francisco, July 14, 2016}

The presentation reviewed Lamport's 1978 paper ``Time, Clocks, and the Ordering of Events in a Distributed System''\citep{lamport1978time}---the second most cited paper in computer science. The talk argued that while Lamport correctly extracted the partial ordering of events from special relativity, he did not extract the \emph{time-symmetry} of the underlying physics.

\marginnote{Video of the talk: \href{https://www.youtube.com/watch?v=CWF3QnfihL4}{YouTube}. Slides: \href{https://speakerdeck.com/pborrill/time-clocks-and-the-reordering-of-events-pwl-san-francisco-14-jul-2016}{SpeakerDeck}. Lamport attended in person.}

Lamport attended the talk in person. His only substantive correction: ``a general theory of concurrency is not my goal'' (regarding slide 69). But his subsequent email exchange revealed a far deeper disagreement about the nature of time itself.

\subsection{The Email Exchange: July--December 2016}

The correspondence between Borrill and Lamport, preserved in full in the companion document \emph{A Conversation with Leslie Lamport on Spacetime}\citep{borrill2016conversation}, crystallized the philosophical divide.

\paragraph{Lamport's Key Insight About His Own Paper}

Lamport offered a remarkable self-assessment that had not previously appeared in the literature:

\begin{quote}
``I have been puzzled by the credit I've received for defining the `happens-before' relation \ldots{} when it seems to be such an obvious corollary of Minkowski's 1908 paper. I've put it down to computer scientists' ignorance of physics \ldots{} I just remembered that the one difference between what I did and what Minkowski did is that he defined causality based on messages that \textbf{could have been sent}, and I defined it based on messages that \textbf{actually were sent}.''
\end{quote}

\marginnote{This distinction---``could have been sent'' versus ``actually were sent''---is precisely the gap that the Category Mistake framework exploits. Lamport's happened-before captures \emph{potential} causality; physics requires \emph{actual} causality.}

This distinction, which Lamport himself considered ``such an obvious modification,'' is in fact the critical design choice that embeds the \fito{} assumption. By restricting attention to messages actually sent, Lamport imposed a direction on information flow that physics does not require.

\paragraph{The Train Station Debate}

The most illuminating exchange concerned a seemingly mundane question: how did Borrill and Lamport both arrive at the Caltrain station at approximately the right time?

Lamport wrote: ``I am amazed by the fact that using different watches that do not communicate with one another, you and I were able to be at the train station nearly simultaneously \ldots{} you might want to contemplate how you define `nearly' in the absence of continuous time.''

\marginnote{Lamport's challenge was precise: if you deny continuous time, how do you explain the practical success of unsynchronized clocks? The answer lies in decoherence and the statistical emergence of macroscopic time.}

Borrill's response identified the mechanism: both watches are immersed in an extraordinarily dense bath of thermodynamic interactions (decoherence events). Like temperature and pressure, macroscopic time \emph{emerges} as a statistical property of extremely large numbers of quantum interactions. Two watches agree to within minutes not because they share a common background of time, but because they are both embedded in a shared thermodynamic environment with a sufficient density of causal interactions to produce convergent counting.

This exchange established two opposing intellectual positions that persist to this day:

\begin{description}[nosep]
    \item[Lamport's position:] Continuous time works well enough in practice; concentrate on the mundane matters where formal reasoning yields significant results.
    \item[Borrill's position:] The assumption of continuous time is the hidden structural error that produces the failures formal reasoning cannot explain---link flapping, iCloud corruption, the FLP impossibility, and the limits of consensus.
\end{description}

\paragraph{The Counting Argument}

Lamport pushed back directly on the claim that time always goes forward: ``I'm quite capable of counting $3, 2, 1, 0, -1, -2, -3, \ldots$''

The response identified the deeper point. Of course one can count backward---the integers are ordered in both directions. But the \emph{physics} of measurement yields only the squared modulus of the probability amplitude (the Born rule), which is always positive. From a local observer's perspective, the arrival of information increments a counter; the departure of information decrements it. The Born rule ensures that what we measure---the \emph{change} we can count---is always non-negative.

\marginnote{The counting argument connects to reversible computing: a non-monotonic count $3, 4, 5, 6, 7, 6, 5, 6, 7, 8 \ldots$ is useful for local recovery from failures in a chain of computations.}

More practically, the response introduced non-monotonic counting as a model for recovery: the sequence $3, 4, 5, 6, 7, 6, 5, 6, 7, 8, 9, \ldots$ represents a computation that advances, encounters a failure, rolls back, and advances again. This is precisely the reversible protocol that OAE implements at the link level.

\section{Stanford EE380: The Datacenter Context (2014, 2016)}

Two Stanford EE380 lectures anchored the ideas in engineering practice.

\subsection{April 2014: Information, Entanglement and Time}

The first lecture\citep{borrill2014stanford} introduced the Subtime Conjecture to a computer engineering audience. The central claim: attempts to construct a ``simultaneity plane'' across datacenters---as Google's Spanner\citep{corbett2013spanner} does with GPS receivers and atomic clocks---are unsound in principle because simultaneity planes are not physically realizable. They can only be approximated statistically, with smaller standard deviations corresponding to higher rates of clock synchronization messages.

\marginnote{The April 2014 talk predates the Google Spanner critique that later appeared in the arXiv papers. The intellectual seeds were planted here.}

At some point, clock synchronization traffic dominates conventional networks and it becomes more economical to use logical clocks and interaction-based serialization foci (two-phase commit or \paxos{}) for distributed transactions. This is because messages carrying explicit causal information are more efficient than messages attempting to maintain a global notion of ``now.''

\subsection{November 2016: The Time-Less Datacenter}

The second lecture\citep{borrill2016ee380}, co-presented with Alan Karp, introduced the engineering realization of these ideas: a network architecture based on direct links between nodes, with liveness maintained by ``hot-potato'' token exchange at the media layer.

The post-session Q\&A\citep{borrill2016stanfordqa} produced several exchanges that previewed the Category Mistake framework:

\begin{description}[nosep]
    \item[On heartbeats:] A conventional heartbeat extends the synchronization domain of the sender. OAE's liveness protocol operates below the protocol stack, using dedicated links with no contention from useful traffic.
    \item[On Google Spanner:] ``We think we've discovered something that goes beyond Google's Spanner and TrueTime \ldots{} Conventional notions of time in computer science cause us to think of time as a \emph{background}. When you dig into the physics, this is a questionable assumption.''
    \item[On the end-to-end argument:] The conventional networking argument---that reliability must be end-to-end because the network is unreliable---holds for switched networks. For direct links with bilateral state, the failure domain is contained, and end-to-end reliability can be composed from link-level guarantees.
    \item[On bandwidth:] The ``off by at most one'' property applies to liveness tokens, not data packets. Application traffic pipelines normally; the link temporarily becomes half-duplex only when atomic mode is invoked.
\end{description}

\section{The Category Mistake Framework (2016--2025)}

The recognition that Lamport's framework embeds a category mistake developed gradually through the analysis of multiple researchers' work.

\subsection{From Lamport to FITO}

The Lamport \fito{} Category Mistake Analysis\citep{borrill2026lamportfito} identifies the specific structural commitments in Lamport's framework:

\begin{enumerate}[nosep]
    \item \textbf{The happened-before relation:} The \fito{} assumption is embedded in the message-passing rule---sending \emph{precedes} receiving, with no mechanism for bilateral causality.
    \item \textbf{Logical clocks:} Clocks only increase. There is no provision for decreasing a timestamp, rolling back to a previous state, or reconciling conflicting causal histories.
    \item \textbf{Concurrent events:} Lamport's arbitrary tie-breaking (by process ID) imposes an order that reflects no physical reality.
    \item \textbf{TLA\textsuperscript{+}:} Behaviors are infinite sequences $s_0, s_1, s_2, \ldots$ with strictly increasing indices. The prime operator refers to the \emph{next} state, never the previous. Past-tense temporal operators, which exist in other logics, are deliberately excluded.
    \item \textbf{\paxos{}:} The word ``eventually'' embeds \fito{}---it assumes time moves forward and that waiting long enough resolves uncertainty.
\end{enumerate}

\marginnote{The category mistake is \emph{not} that Lamport is wrong within his model. It is that the model has been reified---treated as if it captured physical causality rather than computational dependency.}

The key distinction: Lamport's happened-before captures \emph{potential causality within a computation}---what information \emph{could have} flowed through message paths. It does not capture whether information actually flowed, physical causation outside the message system, or the direction of causation (which may be bilateral).

\subsection{The Broader Pattern}

The same \fito{} analysis was applied systematically across distributed systems theory, producing a portfolio of Category Mistake analyses for the foundational figures: Bell, Shannon, Gray, Brewer (CAP theorem), Birman (virtual synchrony), Liskov, Lynch, Stonebraker, Helland, Kingsbury (Jepsen), Herlihy, Rovelli, Smolin, Turing, and von Neumann.\citep{borrill2026catmistake} Each analysis identifies the same structural pattern: the assumption that causality flows forward in time only, embedded at the foundations of the respective framework, producing specific engineering consequences when that assumption fails.

\subsection{What Lamport Got Right}

The Category Mistake framework does not dismiss Lamport's contribution. On the contrary: Lamport's insistence on mathematical rigor---defining ``before'' precisely, providing algorithms for consistent orderings, creating formal tools for specification and verification---is his greatest achievement. Within the \fito{} paradigm, his framework provides rigorous foundations. The critique is that the \emph{paradigm itself} has boundaries, and those boundaries are now being reached at the scales where modern distributed systems operate.

\paxos{} and its variants power the world's most critical systems: Google Spanner, Amazon DynamoDB, Apache ZooKeeper. These systems work because \fito{} assumptions hold well enough at the scales where they operate. The question is what happens at microsecond scales across geographic distances, where the physics of light propagation, link flapping, and network partitioning expose the assumption's limits.

\section{The Practical Consequences (2023--2026)}

\subsection{The 366\,GB Archive}

The theoretical framework found its most dramatic empirical validation in the author's own iCloud Drive.\citep{borrill2026icloud} Over years of normal use across multiple Apple devices, iCloud's sync protocol produced:

\begin{itemize}[nosep]
    \item A 366\,GB ``iCloud Drive (Archive)'' folder containing 110 top-level items that had diverged from the 405\,GB primary iCloud Drive.
    \item 89 folders existing \emph{only} in the archive---files silently removed from the primary store with no notification.
    \item 57 folders with divergent contents between archive and primary.
    \item 1,406 files in the FIGURES directory alone that existed only in the archive.
\end{itemize}

\marginnote{This is what the Category Mistake looks like at operational scale: hundreds of gigabytes of unresolvable divergence, silent deletions, and phantom duplicates requiring weeks of manual effort to repair.}

Resolution required custom Python scripts performing MD5-based comparison of every file in both trees, categorizing conflicts, and merging without data loss. The scripts could not rely on timestamps---iCloud's conflict resolution had already corrupted them---and instead had to treat file content as the only source of truth.

\subsection{The iCloud Failure Taxonomy}

The arXiv paper ``Why iCloud Fails''\citep{borrill2026icloud} documents the structural thesis: iCloud's failures are not random bugs but \emph{projection artifacts} caused by a single structural error:

\begin{quote}
\emph{Projecting a distributed causal graph onto a linear temporal chain destroys essential structure. Information is lost. The losses manifest as corruption, conflicts, stalls, and silent data destruction.}
\end{quote}

The paper identifies five interlocking incompatibilities---Time Machine, git, automated toolchains, the filesystem dialect problem, and link flapping---all arising from \fito{} assumptions embedded in iCloud's sync protocol: timestamp primacy (``later'' means ``more correct''), intermediate state leakage, smash-and-restart recovery, and non-idempotent retry semantics.

\subsection{Empirical Confirmation}

The Waterloo group's comprehensive studies\citep{alquraan2018,alkhatib2023} provided independent empirical confirmation: 136 system failures attributed to network partitioning across 25 widely used distributed systems, with 80\% catastrophic impact, 90\% silent, and 21\% causing permanent damage. iCloud operates in exactly the regime these studies characterize: intermittent connectivity between devices that partition whenever a laptop lid closes or a phone enters a tunnel.

\section{The Engineering Resolution: OAE}

Open Atomic Ethernet represents the constructive half of the argument: a demonstration that protocol semantics exist which avoid the Category Mistake entirely.

\subsection{The Core Principle}

\emph{A transfer is not real until it is known to be real.}

OAE enforces perfect information feedback at the link boundary. Data is only considered delivered once it has been explicitly reflected back to the sender. Knowledge, not elapsed time, defines correctness.\citep{borrill2026oaespec}

\subsection{From Alice-and-Bob to AELink}

The path from the 2016 thought experiment to the 2026 protocol specification is direct. The Alice-and-Bob universe---two entities exchanging a single quantum of information with no background time---is exactly the model that OAE implements at Ethernet Layer~2:

\begin{table}[h]
\small
\begin{tabularx}{\textwidth}{lX}
\toprule
\textbf{Alice-and-Bob (2016)} & \textbf{OAE AELink (2026)} \\
\midrule
Two hydrogen atoms & Two network endpoints \\
Single photon exchanged & Causally significant token \\
No background time & No timeout-and-retry \\
Proper time is zero for the photon & Token transfer is atomic \\
LOV has only two states & Link state machine has known states \\
Observation ``steals'' the photon & Third-party interaction breaks isolation \\
\bottomrule
\end{tabularx}
\end{table}

\marginnote{The conceptual thread from the Simple Time Description through the Lamport conversation to the OAE specification is unbroken. The physics informed the critique; the critique motivated the design.}

\subsection{Bilateral Transactions and Symmetric Reversibility}

OAE links are not passive channels but \emph{joint stateful systems}. Both peers implement identical state machines that evolve synchronously via token exchange.\citep{borrill2026symmetricrev} For every operation, there exists a logically defined inverse that restores the prior state:

\begin{itemize}[nosep]
    \item Partial transactions abort cleanly, returning to equilibrium with no leaked state.
    \item Bit errors can be rolled back without corrupting state.
    \item All token transfers are atomic: they either complete fully or leave the system unchanged.
\end{itemize}

This is the engineering realization of the bilateral causality that Wheeler-Feynman absorber theory describes for physics: a message is not ``sent'' until it is ``absorbed,'' and the absorber's response retroactively constrains the sender's state.

\subsection{The Triangle Topology}

A direct link has exactly one failure hazard: disconnection. Both endpoints know the link is down, but neither can communicate this to the other. The resolution requires a minimum of three nodes connected by direct links---a triangle---such that when any single link fails, the remaining two provide an alternative path.\citep{borrill2026triangle}

\marginnote{The triangle is the minimal causal structure supporting bilateral knowledge under single-link failure. Every triangle is a recovery resource; every link has a designated transaction manager.}

This is precisely the two-phase commit protocol that Jim Gray described: a coordinator proposes, participants vote, and the coordinator resolves. In the triangle topology, each node is simultaneously a participant on two links and a transaction manager for the third. The blocking vulnerability of classical 2PC is eliminated because the coordinator for a failed link is a node \emph{not on that link}---its failure would be a different failure, detected and recovered by a different triangle.

\section{Concept Development Timeline}

\begin{description}
    \item[2013--2014] Physics foundations: background-free time, Subtime Conjecture, Alice-and-Bob thought experiment. Stanford EE380 lecture (April 2014).

    \item[2015] Audio recordings of Lamport-related discussions (February 2015). Continued development of Simple Time framework with exploration of Feynman path integrals, Abramsky's work at Simons Institute.

    \item[2016] \emph{Pivotal year.} Papers We Love talk at GitHub (July 14). Email exchange with Leslie Lamport (July--December). Lamport's self-assessment of happened-before vs.\ Minkowski. Stanford EE380 with Alan Karp (November 16). Simple Time Description and Q\&A documents. Recognition of the ``messages actually sent'' vs.\ ``messages that could have been sent'' distinction as the locus of the category mistake.

    \item[2017--2022] Systematic \fito{} analysis of distributed systems foundations. Development of OAE specification. FITO Category Mistake analyses for Lamport, Bell, Shannon, Gray, Brewer, Birman, Liskov, Lynch, and others.

    \item[2023] iCloud failures become operationally visible: silent folder deletion (April 2023), filename swapping, Keynote image degradation. Discovery of 366\,GB archive divergence.

    \item[2024--2025] OAE specification matures: Pencil and Eraser theorem, symmetric reversibility, triangle topology, exactly-once delivery. Chiplet Summit presentations. Link Wars papers.

    \item[2026] arXiv submissions: ``Why iCloud Fails'' (2602.19433), ``What Distributed Computing Got Wrong.'' Empirical validation through the Lamport file consolidation---a 4,537-file, multi-location corpus requiring SHA-256 checksum verification to resolve iCloud-induced duplication and divergence.
\end{description}

\section{The Unfinished Revolution}

The title of the 2016 Papers We Love presentation---``Lamport's Unfinished Revolution''---names the thesis precisely. Lamport began a revolution in how we reason about distributed systems. He replaced informal reasoning with mathematical rigor. He showed that the partial ordering of events, not the illusion of global time, is the correct foundation.

\marginnote{Lamport took relativity's epistemology---we cannot know the order of distant events---but assumed an ontology: there \emph{is} an order, even if unknown. Physics suggests the ontology may be wrong.}

But the revolution is unfinished. Lamport extracted the partial ordering from relativity while discarding the time-symmetry. He captured computational dependency while assuming physical causality flows forward. He created formal methods that are rigorous within \fito{} while embedding \fito{} so deeply that it became invisible.

The Category Mistake framework completes the revolution by identifying what was assumed, showing where those assumptions fail, and demonstrating---through OAE---that an alternative foundation exists. The alternative does not reject Lamport's work; it extends it beyond the boundaries of \fito{}, into a regime where bilateral causality, symmetric reversibility, and conservation of information replace the forward-only assumptions that physics tells us are incomplete.

The path from two hydrogen atoms exchanging a photon to two Ethernet endpoints exchanging a token is shorter than it appears. It is the same physics, applied at different scales, with the same structural properties. The fifteen years documented in this supplement trace the recognition that this path exists, and the engineering effort to walk it.

\newpage
\appendix

\section{OCP TAP Presentations: A Systematic Deconstruction}\label{app:demolition}

\noindent\textit{The following is the complete text of the ``Demolition'' document, which summarizes the five-part OCP Time Appliances Project presentation series (2024--2025). It is reproduced here verbatim as archival supplementary material.}

\bigskip

\subsection{Episode 1: (What) Clock Coherence From Terrestrial Microdatacenters to Proxima Centauri}

\begin{quote}
\small
Clock synchronization, as we know in this forum, is a hard problem. This talk is motivated by a project Robert G Kennedy III, who revealed a plan to send a swarm of nano-spacecraft to Alpha Centauri and return pictures. [Asilomar Microcomputer Workshop, April 24--26, 2023]. Knowing techniques for distributed clock synchronization is apparently critical to the success of the project he asked us to describe the problems of clock synchronization and illuminate a potential solution based on his work in Distributed Systems in Terrestrial datacenters.
\end{quote}

\noindent\textbf{Video:} \href{https://www.youtube.com/watch?v=3DCpzntiZ_Y}{youtube.com/watch?v=3DCpzntiZ\_Y}\\
\textbf{Slides:} \href{https://daedaelus.com/wp-content/uploads/2024/07/OCP-TAP-Borrill-E1-V1.pdf}{daedaelus.com -- Episode 1 Slides}

\begin{quote}
\footnotesize
There is no now. You cannot synchronize clocks the way you think. Talk originally given at the 2023 Asilomar Microcomputer Workshop, presented live with Jonathan Gorard.

Motivation: (1) to get people thinking about the nature of time and causality, as far removed from the Earth (and TAI/GPS) as possible. (2) to stimulate ``First Principles Thinking'' for Distributed systems.
\end{quote}

\subsection{Episode 2: (Why) A Facade of Newtonianism in Networking}

\begin{quote}
\small
A hidden assumption of Newtonian time pervades all of computer science and networking. We make believe that an ``inertial frame'' allows us to ignore relativity, and that quantum effects are too small to matter. Nothing could be further from the truth. In this session we will explain why this matters: Inconsistent partial orders among different observers (e.g.\ logs in distributed databases), cause silently lost and corrupted data structures. We connect the dots between the hidden assumptions (smooth, monotonic irreversible), and the evidence/consequences seen in recent work on Partial Network Partitioning (PNP).
\end{quote}

\noindent\textbf{Video:} \href{https://www.youtube.com/watch?v=QSJAEjFbqas}{youtube.com/watch?v=QSJAEjFbqas}\\
\textbf{Slides:} \href{https://daedaelus.com/wp-content/uploads/2024/07/OCP-TAP-Borrill-E2-V1-FInal.pdf}{daedaelus.com -- Episode 2 Slides}

\marginnote{Key claims from Episode 2: Clock synchronization error is indistinguishable from latency. Irreversibility and ``causal order'' are in the eye of the observer---not guaranteed to be consistent across different observers.}

\begin{quote}
\footnotesize
When we think about clocks as an incrementing number, we are committing the FITO fallacy---``Forward In-Time-Only'' Thinking:
\begin{itemize}[nosep]
\item Counterfactuals---``events that could have occurred but eventually did not, play a unique role in quantum mechanics in that they exert causal effects despite their non-occurrence''
\item Clock Synchronization Error is indistinguishable from Latency
\item Irreversibility (Monotonicity) is an illusion not guaranteed by physics, unless we build Ancilla to explicitly manipulate causality
\item Irreversibility and ``causal order'' are IN THE EYE OF THE OBSERVER---not guaranteed to be consistent across different observers
\end{itemize}
\end{quote}

\subsection{Episode 3: Race Conditions and Exactly Once Semantics in Distributed Systems}

\begin{quote}
\small
In Episode 1 (What) and Episode 2 (Why) we showed how misunderstandings accumulate within a Newtonian framework of time, and how this leads to lost transactions and corrupted data. In this Episode we help the audience make the leap from Newtonian Time (what we know for certain that just ain't so) to Post-Newtonian Time (relativistic SR/GR, and Indefinite Causal Order (ICO). Now PTP is widely available, we propose experiments to falsify beliefs about the Newtonian Time. We will then show how to utilize the excellent engineering behind PTP and PTM, from a different perspective: using the clock hierarchy to build causal trees with reliable failover, to help address race conditions and achieve Exactly Once Semantics.
\end{quote}

\noindent\textbf{Video:} \href{https://www.youtube.com/watch?v=ll5aF-oN-YE}{youtube.com/watch?v=ll5aF-oN-YE}\\
\textbf{Slides:} \href{https://daedaelus.com/wp-content/uploads/2024/07/OCP-TAP-Borrill-Episode3-DARK.pdf}{daedaelus.com -- Episode 3 Slides}

\subsection{Episode 4: Timeout and Retry (TAR) in Distributed Systems}

\begin{quote}
\small
Why we can't have nice things in Distributed Systems.
\end{quote}

\noindent\textbf{Video:} \href{https://www.youtube.com/watch?v=ll5aF-oN-YE}{youtube.com/watch?v=ll5aF-oN-YE}\\
\textbf{Slides:} \href{https://drive.google.com/file/d/1yXLCt8uCUVhh-pBttNlqcYocArvN_Yye/view?usp=drive_link}{Google Drive -- Episode 4 Slides}

\marginnote{Key claims from Episode 4: Instants are meaningless, only intervals (on the same computer/timeline) are relevant. Photons don't carry timestamps, but timestamps are carried by photons. Shannon entropy is a logarithm; the logarithm of zero is minus infinity.}

\begin{quote}
\footnotesize
\begin{itemize}[nosep]
\item Instants are meaningless, only intervals (on the same computer/timeline) are relevant
\item Photons don't carry timestamps, but timestamps are carried by photons
\item The speed of light is the \emph{pivot} around which time and space evolve
\item Timeout and retry (TAR) on different timelines will silently corrupt data structures
\item Shannon entropy is a logarithm. The logarithm of zero (no information) is minus infinity.
\item Bayesian approaches require a prior belief, which can be unbounded (zero to infinity). Actually, it's much worse: can be $-\infty, -1, -0, +0, +1, +\infty$. We can't do Bayesian statistics under those conditions, mathematically, their results are undefined
\item Shannon Entropy is uncertainty, and the same problem applies when you apply the set $\{-\infty, -1, -0, +0, +1, +\infty\}$ to Information and Entropy $p\log(p)$
\item Measurements \emph{appear} instantaneous because there is no background of time on which to measure anything. Timestamps don't provide causal order.
\end{itemize}
\end{quote}

\subsection{Episode 5: Ethernet Spacetime}

\begin{quote}
\small
Timestamps are an Illusion. They can't be fixed in software. The quest for a single, consistent timeline across distributed systems collides with the reality that physics itself does not provide a universal notion of time---and in many quantum scenarios, there isn't even a consistent causal order at all. Though timestamps will remain indispensable in engineering practice, we must recognize them as approximations rather than absolutes, and design our systems accordingly. Time synchronization is not merely a technical challenge---it's a conceptual one. Software Fixes Are Only Band-Aids, they can never be deterministic. In this presentation, we also announce the birth of Ethernet 2025.
\end{quote}

\noindent\textbf{Video:} \href{https://www.youtube.com/watch?v=kJMdgaYW66k}{youtube.com/watch?v=kJMdgaYW66k}\\
\textbf{Slides:} \href{https://daedaelus.com/wp-content/uploads/2025/01/OCP-TAP-Jan-1-Ethernet-2025-Slides.pdf}{daedaelus.com -- Episode 5 / Ethernet 2025 Slides}

\marginnote{Episode 5 (January 1, 2025) launched the Ethernet-2025 Project, which became the OCP Open Atomic Ethernet (OAE) Project.}

\begin{quote}
\footnotesize
THERE IS NO GLOBAL DRUM BEAT. In Episodes 1 through 4 we expressed doubts about the common belief system of a Newtonian view of the world in this community. We showed how to think about race conditions, and why Timeouts and Retries (TAR) are the root of all evil. Our conclusion is that Timestamps are an Illusion. They can't be fixed by software.

The quest for a single, consistent timeline across distributed systems collides with the reality that physics itself does not provide a universal notion of time---and in quantum mechanics (the machine code of our universe), there is no consistent causal order at all. We cannot, therefore, rely on this illusion of an irreversible drumbeat on an inaccessible ``real line'' to provide linear time order for events in our networked systems.

Although timestamps will remain indispensable in engineering practice, we must recognize them as approximations rather than absolutes, and design our systems accordingly.
\end{quote}

\subsection{Papers We Love: Time, Clocks and the Reordering of Events}

\noindent\textbf{Video:} \href{https://www.youtube.com/watch?v=CWF3QnfihL4&t=4552s}{youtube.com/watch?v=CWF3QnfihL4}\\
\textbf{Slides:} \href{https://files.speakerdeck.com/presentations/fc5a002e9a4e419b83f0ac5b75c869a9/Lamport_PWL-14-Jul-2016a.pdf}{SpeakerDeck -- Lamport PWL Slides}

\begin{quote}
\small
Leslie Lamport's original paper 1978 paper, \textit{Time, Clocks, and the Ordering of Events in a Distributed System}, introduced important concepts such as logical clocks and the notion of ``happened before'', both of which are key ideas in understanding causality in distributed systems. Lamport's paper laid the groundwork that allows us to bypass the need for perfect synchronization in physical clocks, which is known to near impossible. However, due to a continuous misunderstanding of the nature of time in both physical and computer science, his work remains largely unfinished. In this talk, Paul highlights how traditional synchronizing methods such as PTP and NTP are fundamentally flawed because they assume a global notion of time through which perfect simultaneity can be achieved. To combat this problem, Paul introduces reversible computing, through which nodes in a system can undo events to recover from errors. Granting individual nodes with this ability reduces the need for a global fallback system and allows for nodes to have greater autonomy across their local neighborhoods.

Paul emphasizes that our conventional notions of time being linear, uniform, and discrete are misleading and make assumptions about the physical nature of our universe that simply don't exist. Rather, time should be thought about in the form of causal relationships that branch off from one another in a tree like structure. However, disagreements and discourse between physicists, mathematicians, and computer scientists limit progress in this regard. In order to solve problems of time and concurrency in distributed systems, increased collaboration is required.
\end{quote}

\section{Review: D{\AE}D{\AE}LUS OCP TAP Presentations, 2024--2025}\label{app:review}

\begin{quote}
\small\noindent
Through a systematic five-part series presented to the Open Compute Project's Time Appliances Project, Paul Borrill has demonstrated that the foundational assumptions of distributed systems---particularly the ACID properties of database transactions---rest upon a \textit{fa\c{c}ade of Newtonianism} that is fundamentally incompatible with 20th Century physics. This analysis examines how modern understanding of relativity, quantum mechanics, and information theory renders impossible the very concepts upon which computer science has built its theoretical framework.
\end{quote}

\subsection{The Collapse of Simultaneity}

The conventional wisdom in distributed systems assumes the existence of a global timeline---what Borrill calls the ``global drum beat.'' This assumption underlies every attempt to create consistent distributed state, from Google's Spanner to AWS's microsecond guarantees to the Time Appliances Project itself.

\marginnote{Einstein demonstrated over a century ago that simultaneity is observer-dependent. The fundamental premise of distributed systems---that we can establish a consistent global ordering of events---is physically impossible.}

Yet Einstein demonstrated over a century ago that simultaneity is observer-dependent. The fundamental premise of distributed systems---that we can establish a consistent global ordering of events---is \textit{physically impossible}.

What we call ``clock synchronization'' is actually the construction of causal trees, not the establishment of universal time. Every timing protocol creates a hierarchy of causality, not a plane of simultaneity.

\subsection{The FITO Fallacy and Temporal Irreversibility}

Modern distributed systems suffer from what Borrill terms the ``FITO fallacy''---Forward In-Time-Only thinking. This assumption of monotonic time progression underlies the Atomicity and Durability properties of ACID transactions.

However, irreversibility itself is observer-dependent. What appears irreversible to one observer may not be to another, particularly when quantum mechanical effects are considered.

The assumption of monotonicity---that timestamps always increase---is an illusion not guaranteed by physics. Without external mechanisms (ancilla) to enforce causal order, distributed systems cannot guarantee the temporal ordering upon which ACID properties depend.

\subsection{The Information-Theoretic Impossibility}

The problems extend beyond relativity into fundamental information theory. Borrill demonstrates that the timeout and retry (TAR) mechanisms essential to distributed systems create mathematical impossibilities when applied across different timelines.

The measurement problem is more fundamental than engineering approximations suggest: ``instants are meaningless, only intervals (on the same computer/timeline) are relevant.'' This directly contradicts the notion of atomic transactions, which require well-defined instants at which operations either occur or do not occur.

Moreover, the information propagation constraint creates an indistinguishability between clock synchronization error and network latency. Since ``photons don't carry timestamps, but timestamps are carried by photons,'' the very act of timestamp distribution introduces uncertainty that cannot be eliminated.

\subsection{Shannon Entropy and the Bayesian Breakdown}

The mathematical foundations supporting distributed consensus suffer from fundamental problems in information theory. Shannon entropy, defined as $H = -\sum p_i \log p_i$, becomes undefined when probabilities approach zero.

Bayesian approaches to distributed consensus require prior beliefs, but in distributed systems these priors can range over the set
\begin{equation}
\{-\infty, -1, -0, +0, +1, +\infty\}.
\end{equation}
Under these conditions, Bayesian inference becomes mathematically undefined, undermining probabilistic approaches to achieving consistency.

\subsection{Quantum Mechanical Causal Indefiniteness}

The final blow to distributed systems foundations comes from quantum mechanics. Modern physics recognizes indefinite causal order (ICO) as fundamental. In quantum mechanics, ``there is no consistent causal order at all.''

Perhaps most remarkably, quantum counterfactuals---``events that could have occurred but eventually did not''---can ``exert causal effects despite their non-occurrence.'' This suggests that distributed systems must account for the causal influence of operations that never actually execute.

\subsection{The Systematic Failure of ACID Properties}

Borrill's analysis reveals how each ACID property fails under physical scrutiny:

\textbf{Atomicity} requires well-defined instants where operations occur or do not occur. But physics tells us that ``measurements `appear' instantaneous because there is no background of time on which to measure anything.'' The atomic instant is a mathematical fiction.

\textbf{Consistency} assumes a global state that can be maintained across distributed nodes. But consistency requires simultaneity, which ``doesn't exist (except in an empty frozen universe).'' Global consistency is physically impossible.

\textbf{Isolation} assumes operations can be segregated from one another temporally. But when ``timeout and retry (TAR) on different timelines will silently corrupt data structures,'' true isolation becomes impossible across distributed timelines.

\textbf{Durability} assumes irreversible state changes. But ``irreversibility is an illusion not guaranteed by physics'' and is fundamentally ``in the eye of the observer.'' What appears durable to one observer may not be to another.

\subsection{The Engineering Illusion}

Borrill acknowledges that ``timestamps will remain indispensable in engineering practice'' but insists we must ``recognize them as approximations rather than absolutes.''

The Time Appliances Project, Google Spanner, and AWS microsecond guarantees represent sophisticated engineering attempts to create what amounts to a ``simultaneity plane''---but this remains fundamentally an illusion. These systems work not because they solve the physical impossibilities, but because they create sufficiently convincing approximations within bounded error tolerances.

\subsection{Implications for Distributed Systems Design}

The recognition that distributed systems operate on physically impossible foundations suggests a fundamental reorientation is needed. Rather than attempting to create global consistency through temporal synchronization, systems should embrace causal ordering and accept the observer-dependent nature of temporal relationships.

Borrill's proposed solution involves using timing hierarchies to construct causal trees rather than attempting clock synchronization. This represents a shift from fighting physics to working with it.

The quest for ``exactly once semantics'' and perfect consistency may be as futile as medieval attempts to build perpetual motion machines---not because the engineering is insufficient, but because the fundamental premises violate physical law.

\subsection{Conclusion}

Paul Borrill's systematic deconstruction reveals that computer science has been built upon a fa\c{c}ade of Newtonianism that modern physics has thoroughly demolished. The ACID properties, the foundation of reliable distributed systems, assume a universe that does not exist---one with absolute time, universal simultaneity, and observer-independent causality.

This does not mean distributed systems cannot function, but it does mean we must abandon the illusion that we can engineer our way around fundamental physical constraints. Instead, we must design systems that acknowledge and work with the observer-dependent, probabilistic, and causally indefinite nature of reality itself.

The Time Appliances Project and similar efforts represent humanity's attempt to impose Newtonian order on a quantum mechanical universe. They succeed as engineering approximations but fail as fundamental solutions to the problems they purport to solve. Recognition of this distinction may be the first step toward building distributed systems that work with physics rather than against it.

\bibliography{lamport_background_refs}
\bibliographystyle{ieeetr}

\end{document}